\documentclass[pre,twocolumn]{revtex4}
\pdfoutput=1
\usepackage{graphicx}
\usepackage{dcolumn,amsmath,amssymb}
\usepackage{bm}

\begin{document}

\title{Continuous wavelet transform based time-scale and multi-fractal analysis of the nonlinear oscillations in a
hollow cathode  glow discharge plasma}
\author{Md. Nurujjaman}
\email{jaman_nonlinear@yahoo.co.in} 
\altaffiliation[Present address: ]{TIFR Centre For Applicable Mathematics, Sharda Nagar, Chikkabommasandra, Bangalore
- 560065, India.}
\affiliation{Plasma Physics Division, Saha Institute of Nuclear Physics, 1/AF, Bidhannagar, Kolkata -700064, India}

\author{Ramesh Narayanan}
 \email{rams@plasma.inpe.br}
 \altaffiliation[Present address: ]{Laboratorio Associado de Plasma, Instituto Nacional de Pesquisas Espaciais,
 Av. dos Astronautas, 1758 - Jardim da Granja 12227-010 Sao Jose dos Campos, SP, Brazil}
\affiliation{Plasma Physics Division, Saha Institute of Nuclear Physics, 1/AF, Bidhannagar, Kolkata -700064, India}

\author{A.N. Sekar Iyengar}
\email{ansekar.iyengar@saha.ac.in}

\affiliation{Plasma Physics Division, Saha Institute of Nuclear Physics, 1/AF,
Bidhannagar, Kolkata -700064, India}

\begin{abstract}

Continuous wavelet transform (CWT) based time-scale and
multi-fractal analyses have been carried out on the anode glow
related nonlinear floating potential fluctuations in  a hollow
cathode glow discharge plasma. CWT has been used to obtain the contour
and ridge plots. Scale shift (or inversely frequency shift) which is a
typical  nonlinear behaviour, has been detected from the
undulating  contours. From the ridge plots, we have identified the
presence of nonlinearity and degree of chaoticity. Using the wavelet transform modulus maxima 
technique we have obtained the multi-fractal spectrum for the
fluctuations at different discharge voltages and the spectrum was
observed to become a monofractal for periodic signals. These
multi-fractal spectra were also used to estimate different
quantities like the correlation and fractal dimension, degree of
multi-fractality and complexity parameters. These estimations have
been  found to be consistent with the nonlinear time series
analysis.
\end{abstract}

\maketitle

\section{Introduction}
Glow discharge plasma is a typical complex medium exhibiting a
wide variety of nonlinear phenomena such as chaos, noise induced
resonances, self organized criticality, frequency entrainment,
complex structures, etc.~\cite{Chaos:jaman,pre:jaman,pla:jaman,prl:Lin
I,pop:dinklage,pre:Klinger,prl:lashinsky}, which have been studied
using a wide variety of analysis methods, such as the nonlinear time
series analysis, spectral analysis, etc. Nonlinear time series
analysis like the estimation of the correlation dimension,
fractal dimension and Lyapunov exponents reflect only the
asymptotic and global behavior of the system but cannot identify the degree of chaoticity, local structures, etc. These estimations are also too simplistic to characterize a time
series, because the estimation emphasizes regions of the attractor
of a time series which is frequently visited by its orbit in the
phase
space~\cite{book:Sprott,IJBC:Sprott,pre:small,physicaD:Chandre,book:Kantz}.
Further, the estimation  of these exponents require very large
data sets, and degrade rapidly with noise
~\cite{book:Sprott,physicaD:gilmore}.

Spectral analysis (based on Fourier techniques)  has also been
used  extensively to identify the different routes to chaos, frequency
entrainment, and many other nonlinear phenomena in plasma
~\cite{Chaos:jaman,pre:Klinger,prl:lashinsky,ppcf:boswell}.
However, the spectral analysis has a limitation of yielding only  
the spectral amplitude,  neglecting the phase  information,  as a result of which
 it can not distinguish between any two time series
of the same spectral amplitude~\cite{pre:small}. Since the short time scale
characteristics of a signal  are also overlooked, information on 
the time history and the local properties like singularity
exponents and the scale shift (or inversely frequency
shift) are lost~\cite{pre:small, PhysicaD:Higuchi,physicaD:Chandre}. Hence one has
to resort to  a time-scale based analysis consistent with
the nonlinear time series analysis, that can detect the presence of
chaos, nonlinearity and other local structures in the
system and for this purpose continuous wavelet transform (CWT) is one of the best candidates.

CWT can be used to construct contours, which in turn are useful to
study some of the nonlinear properties like the scale shift as well as periodic
behaviour. Characterization of the weak and strong chaos is possible
by ridge plots constructed from the CWT decomposition of a chaotic
time series~\cite{physicaD:Chandre}. Statistical scaling
properties of a complex signals can be characterized by the CWT based
multi-fractal spectra and these spectra can also be used to
estimate several exponents like the fractal dimension, correlation
dimension, etc., that are useful to characterize nonlinear and
chaotic nature of a time
series~\cite{PhysicaD:arneodo,pre:silchenko,pra:chhabra,physicaA:struzik,prl:muzy,pre:muzy,book:mallat,prb:Grussbach}.

In this paper, we have reported some results of the analysis of the  floating
potential fluctuations in a dc glow discharge plasma using
contour and ridge plots based on the CWT decomposition. We have
also constructed the multi-fractal spectra using the wavelet transform
modulus maxima and from these spectra,
the fractal dimension, correlation dimension and other complexity
parameters have been estimated.   The rest of the article is
organized as follows: we present a brief description of the
experimental setup and results in Sections ~\ref{section:Setup}
and ~\ref{sec:results} respectively. Analysis of these results
have been presented in Sections~\ref{sec:cwt},~\ref{sec:ridge} and
\ref{sec:multifractal} respectively. Finally, we discuss and
conclude the results in Section~\ref{sec:conclusion}.

\section{Experimental Setup}

\label{section:Setup}

\begin{figure}
\center
\includegraphics[width=8.5cm]{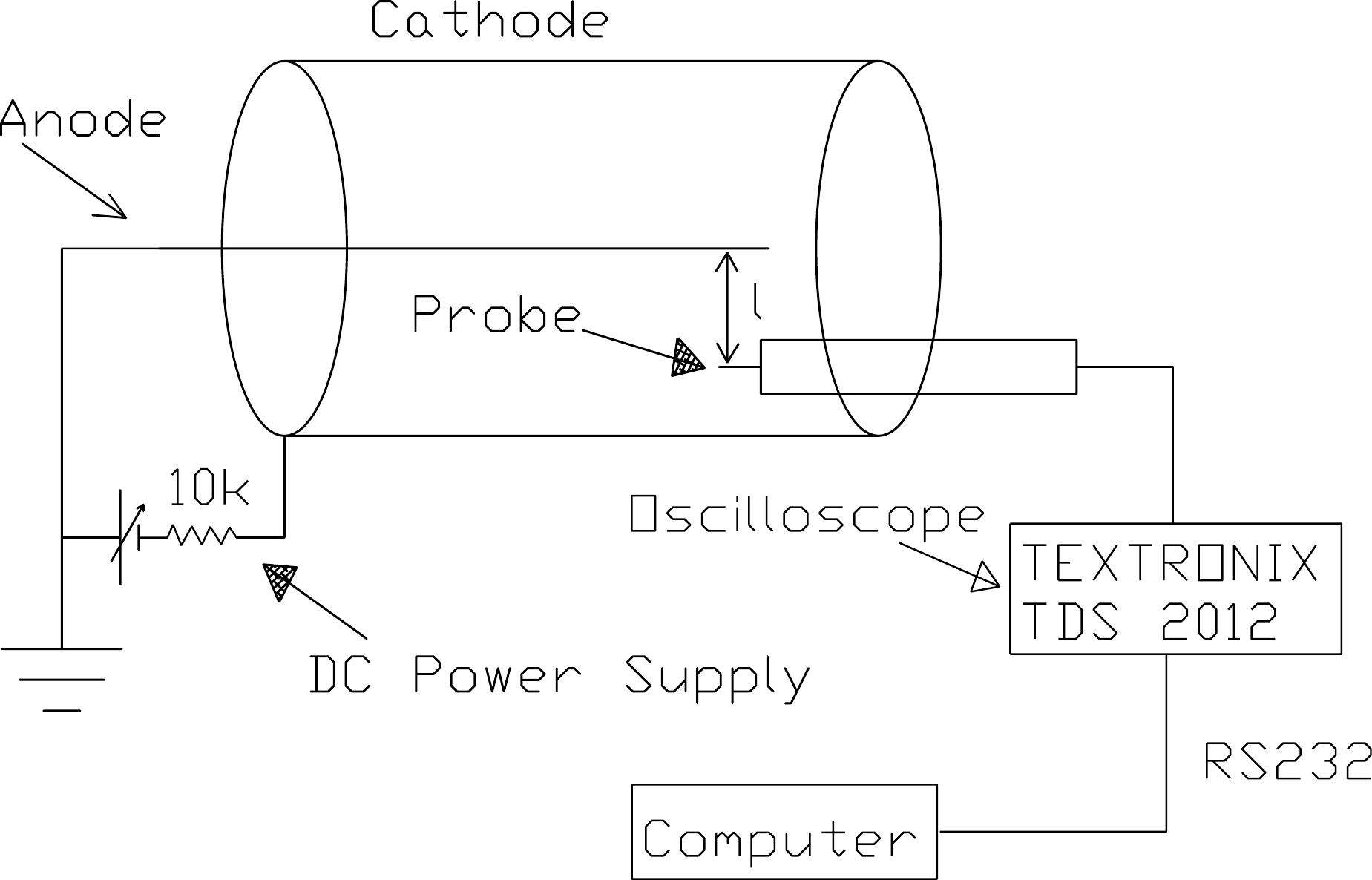}
\caption{Schematic diagram of the hollow cathode discharge tube.
Cathode is a cylindrical hollow tube and the anode is a rod of  stainless
steel.}

\label{fig:setup}
\end{figure}

 The experiments were performed in a hollow cathode dc argon discharge plasma~\cite{Chaos:jaman,pre:jaman,pla:jaman} whose schematic diagram
is shown in Fig~\ref{fig:setup}. The hollow stainless steel tube of diameter 4 cm was the cathode and the
central rod of diameter 2 mm was the anode. The whole assembly was
mounted inside a vacuum chamber and was pumped down to 0.001 mbar.
Then it was filled with argon gas up to 0.95 m bar and a discharge was
struck by a dc discharge voltage. The main diagnostics was the
Langmuir probe used to monitor the floating potential
fluctuations. The typical plasma density was about $10^7cm^{-3}$ and the
electron temperature estimated was about $3-4$ eV.

\section{fluctuation results}
\label{sec:results}
\begin{figure*}
\center
\includegraphics[width=14 cm]{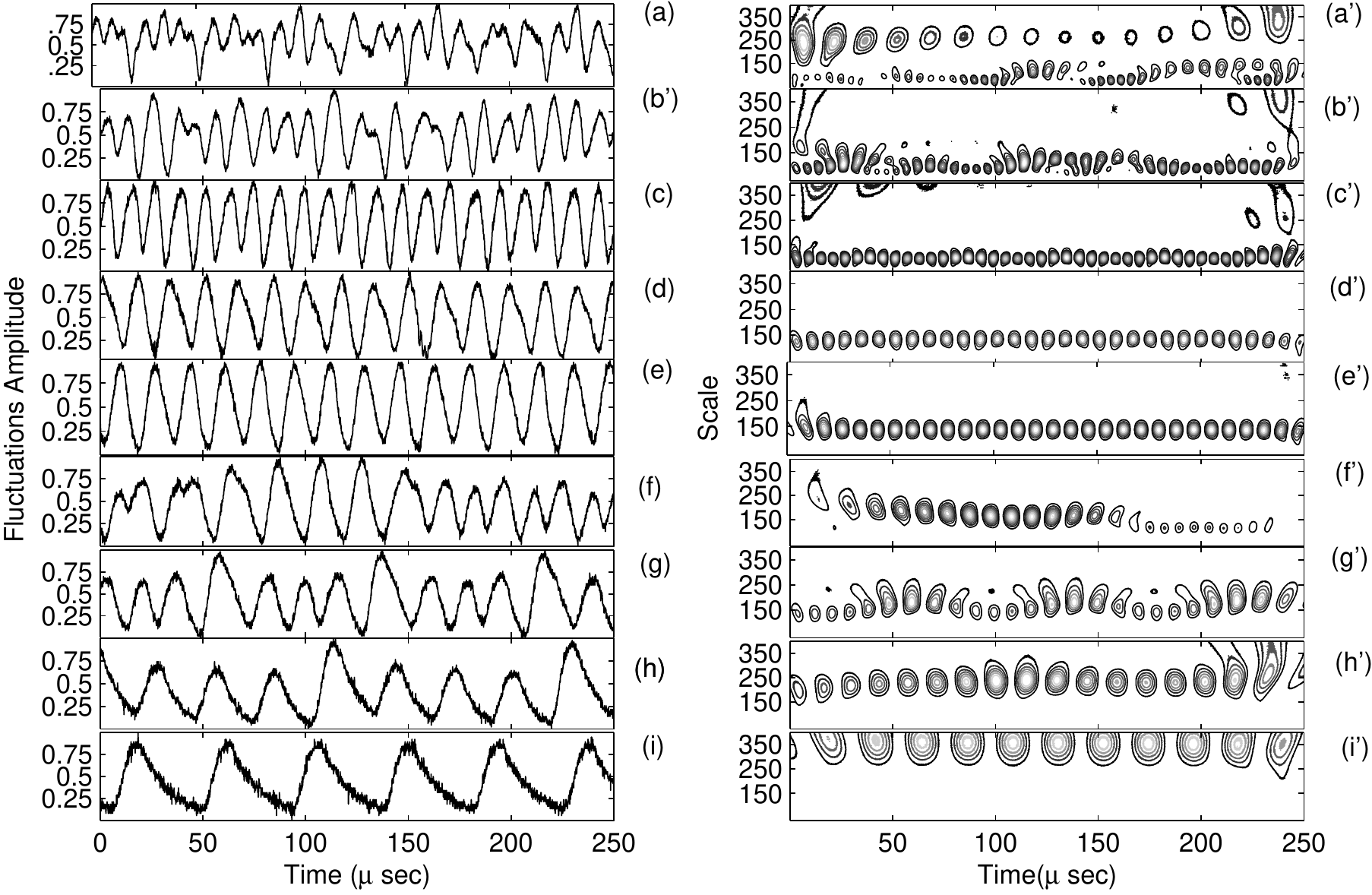}
\caption{(Left panel) Floating potential fluctuations (a$-$i) and (right panel) their
corresponding contour plots of continuous wavelet transforms (a$'-$i$'$) are
shown at 0.95 m bar for different voltages: (a) 283 V; (b) 284 V;
(c) 286 V; (d) 288 V; (e) 289 V; (f) 290 V; (g) 291 V; (h) 292 V;
(i) 293 V. Nonlinear scale shifts are clearly seen in
a$'$ and b$'$. c$'-$e$'$ is almost periodic and with further increase in
DV the signals exhibit nonlinear scale shift before becoming
periodic relaxation oscillations (h$'-$i$'$).} \label{fig2:contour}
\end{figure*}

 The formation of the plasma depends on two main parameters: the discharge voltage (DV) and
the neutral gas pressure. An anode glow was formed around the
anode when a discharge was struck above a threshold neutral
pressure and DV. It was observed that at the initial stage of the
discharge, the anode glow was large in size and decreased with
increase in DV, until it got extinguished beyond  a certain
DV~\cite{Chaos:jaman}. At the instant of formation of the
anode glow, the associated fluctuations in the floating potential
were chaotic which became quasi-periodic with increase in the DV
and finally vanished to a fixed point when the anode glow
disappeared. At this time the plasma inside the hollow cathode became
brighter and filled up the entire plasma column. Details of the
anode glow and the associated instabilities have been discussed in
Ref~\cite{Chaos:jaman}. For the present paper, the floating
potential fluctuations were obtained keeping   pressure constant
at 0.95 m bar. At this pressure the discharge was initiated at
$\approx 283$ V and simultaneously irregular oscillations in the
floating potential were observed [Fig~\ref{fig2:contour}(a) (left
panel)]. With increasing DV, the regularity of the oscillations
increases and final form of the fluctuations just before their
disappearance was the relaxation type of oscillation as shown in
Fig~\ref{fig2:contour}(h$-$i)[left panel].

\section{CWT analysis of the fluctuations}
\label{sec:cwt}
The continuous wavelet transform (CWT) is used to decompose a signal using wavelets, i.e.,
into small oscillations that are highly localized in time and for a signal $\phi(t)$ it
is defined as~\cite{book:mallat}:
\begin{equation}
\label{eqn:wt}
W_\Psi(a,\tau)=\int \phi(t)\Psi_a(t-\tau)dt
\end{equation}
where, $\phi(t)$ is the signal and $\Psi_a(t)$ is an oscillating
function that decays rapidly with time and is termed as
wavelet. $a$ and $\tau$ are the scale and temporal propagation parameters
respectively. Scale `$a$' can be considered to be the  inverse of frequency and so one can estimate the 
frequency  ($F_a$)  corresponding to a scale `$a$' for a particular
wavelet using the relation
$F_a=\frac{F_c}{a\Delta}$~\cite{web:scal2frq}, where $F_c$ is
the center frequency of the analyzing wavelet  and $\Delta$ is the sampling
period of the signal. So a signal can be decomposed in time-scale plane or
time-frequency plane using above scale-frequency relation.  In this
paper, we have used the time-scale decomposition using the
Daubechies orthogonal wavelet  to detect the singularities present in
the time series as it has a slight asymmetric structure, uses few
coefficients and is a good representation of low-order
polynomials~\cite{elsevier:mendes,elsevier:oliveira}.
Fig~\ref{fig2:contour}(a$'-$i$'$)  shows the contours of
$W_\Psi^2(a,\tau)$ for  the signals shown in Figs~\ref{fig2:contour}(a$-$i), in
the time-scale plane. Though one can see that the oscillations go
through different types of behaviour with increasing DV, changes are much more clearer in the CWT contour plots shown in
Fig~\ref{fig2:contour} (right panel). Undulating nature of the
contours in Fig~\ref{fig2:contour}(a$'-$b$'$)
and Fig~\ref{fig2:contour}(f$'-$g$'$) is due to the scale  shift with
time which is one of the signatures of the presence
of nonlinearity in the system~\cite{mssp:lind}. Periodic behaviour
of the fluctuating oscillations of Fig~\ref{fig2:contour}(c$-$e)
is also prominent from the periodic  contours
[Fig~\ref{fig2:contour}(c$'-$e$'$)]. Though the contour plots of the
relaxation oscillations [Fig~\ref{fig2:contour} (h$-$i)] have no
such distinct structures, still one can find differences from the
other two cases. Importance of the CWT based contour plots is
that they can be used to identify the nonlinear nature of a
system, and this identification is very essential for the nonlinear time series analysis.
The CWT coefficients ($W_\Psi(a,\tau)$) are also useful to devise
another plot termed as a ridge plot~\cite{physicaD:Chandre}
which can be exploited to analyse chaotic data, and this  has been
presented in the next section.

\section{ridge analysis}
\label{sec:ridge}

\begin{figure*}
\center
\includegraphics[width=13cm]{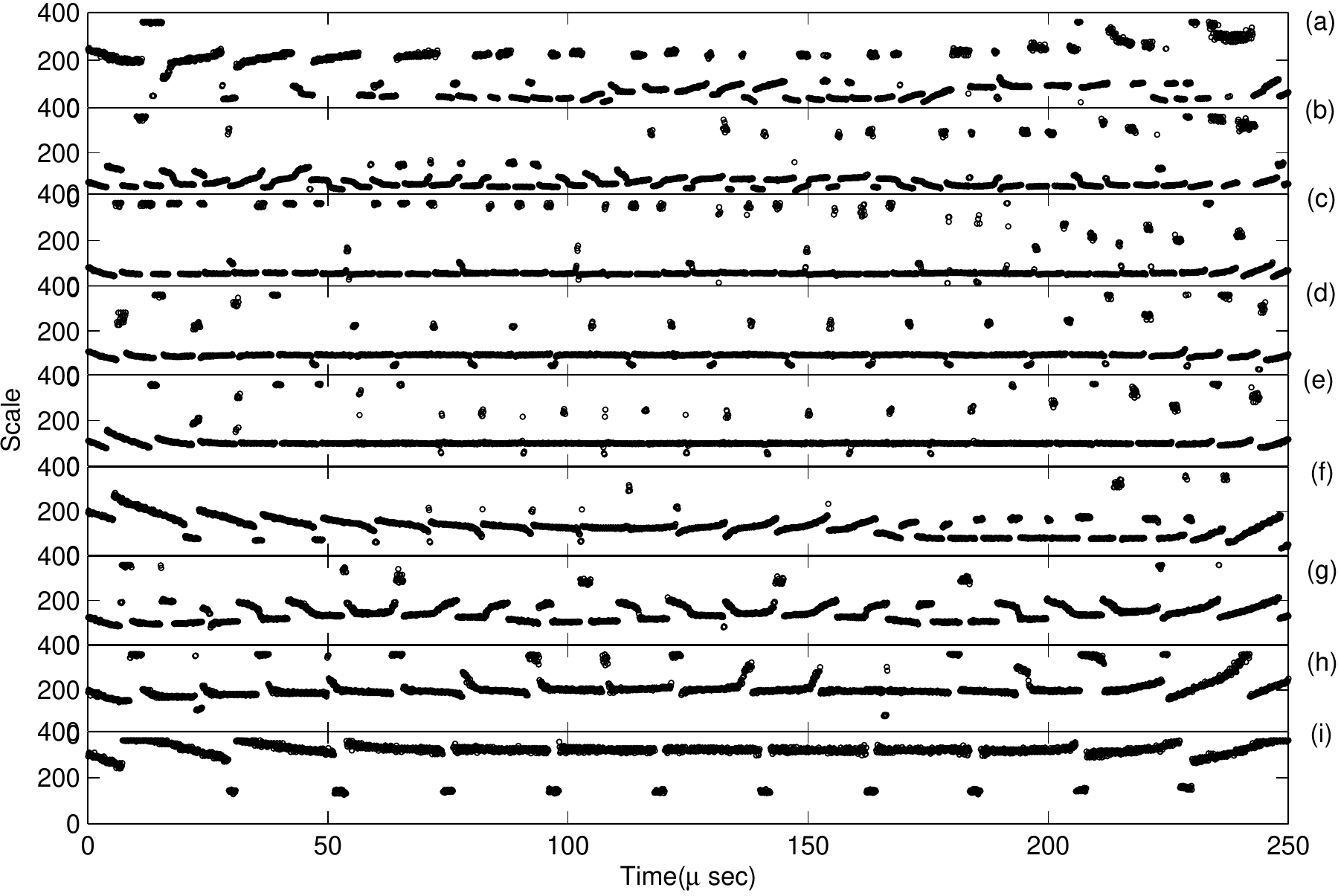}
\caption{Ridge plots along the scales of  the  contour plots shown
in  Fig~\ref{fig2:contour} (right panel).} \label{fig:ridges1}
\end{figure*}

The line joining  the maximum of the CWT coefficients along the
scales `$a$' is called a ridge plot which
can be used to detect the presence of chaos and degree of
chaoticity or periodicity in a system~\cite{physicaD:Chandre}.
Here, we have considered only the prominent ridges to construct the
plots. In Fig~\ref{fig:ridges1}(a) the presence of the broken  ridges  at
the higher scale of 200 and also at the lower scale of about 30
indicates that at the initial stage of DV (283 V) the system is
very chaotic~\cite{physicaD:Chandre}. At 284 V
[Fig~\ref{fig:ridges1}(b)] the system is still  chaotic but the
discontinuous ridges are seen only  at the lower scales and chaos
is not so strong and these results are consistent with the nonlinear time
series analysis~\cite{Chaos:jaman}. With increase in the DV, the
fluctuations [Fig~\ref{fig:ridges1}($c-e$)] became almost regular
and their ridges  are almost constant with time.  The ridge plots
of the fluctuations shown in Figs~\ref{fig2:contour}(f) and (g)
also show broken ridges  indicating reappearance of the chaos.
Appearance of a chaotic window after the periodic case is always
possible and is consistent with previous analysis presented in
Ref~\cite{Chaos:jaman}. Fig~\ref{fig:ridges1}(h) shows
discontinuous ridges but of slightly longer durations and finally
Fig~\ref{fig:ridges1}(i) shows a continuous ridge indicating
periodic signals with breaks  at the fall phase of the relaxation
oscillations at the lower scale of about 300. Here it also shifts
between two scales one at the top and the other at the bottom
indicating the nonlinear frequency shift.

\section{Multifractal spectrum}
\label{sec:multifractal}

In order to understand the scaling properties of the signals we
have constructed multi-fractal spectrum using the wavelet coefficients
of the CWT decomposition. The multi-fractal spectrum ($D(\alpha)$) of
a signal can be estimated using the wavelet transform modulus
maxima method(WTMM)
~\cite{prl:muzy,pre:muzy,book:mallat,chaos:Ivanov} as follows:

\begin{figure}
\center
\includegraphics[width=8.5cm]{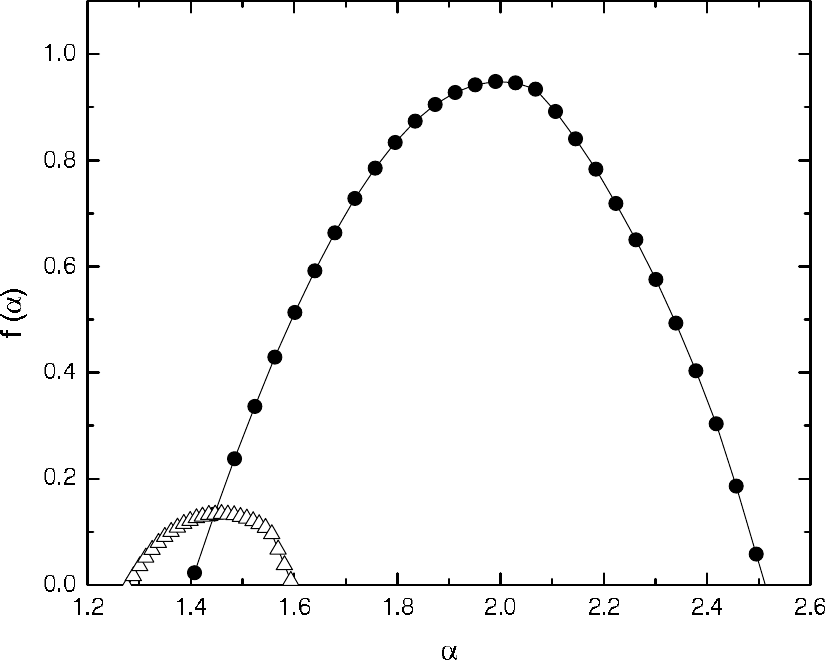}
\caption{$-\bullet-$ shows the multi-fractal spectra at the
initial stage of discharge (283 V). $-\triangle-$ at 292 V,
i.e., just before the steady fixed point is reached and mono fractal in nature.}
\label{fig:mult}
\end{figure}

The partition function $Z(a,q)$ of moment q for a particular scale
`$a$' is defined as a sum of the wavelet transform modulus maxima at that
particular scale is given as
\begin{equation}
Z(a,q)=\sum_{\{t_i(a)\}_i} |W_{\Psi}(a,t_i(a))|^q\sim a^{\tau(q)}
\end{equation}
where, $W_{\Psi}(a,t_i(a))$ is the wavelet modulus maxima and $q$ is the moment considered.
From $\tau(q)$ we can estimate the singularity spectrum using following relations~\cite{book:mallat}

\begin{subequations}
\label{allequations} 
\begin{eqnarray}
\alpha&=&\frac{d\tau(q)}{dq}-\frac{1}{2},\label{equationa}
\\
D(\alpha)&=&\min_{q\in \Re}(q((\alpha +1/2)-\tau(q)),\label{equationb}
\end{eqnarray}
\end{subequations}

\begin{figure*}
\center
\includegraphics[width=12.5cm,height=8cm]{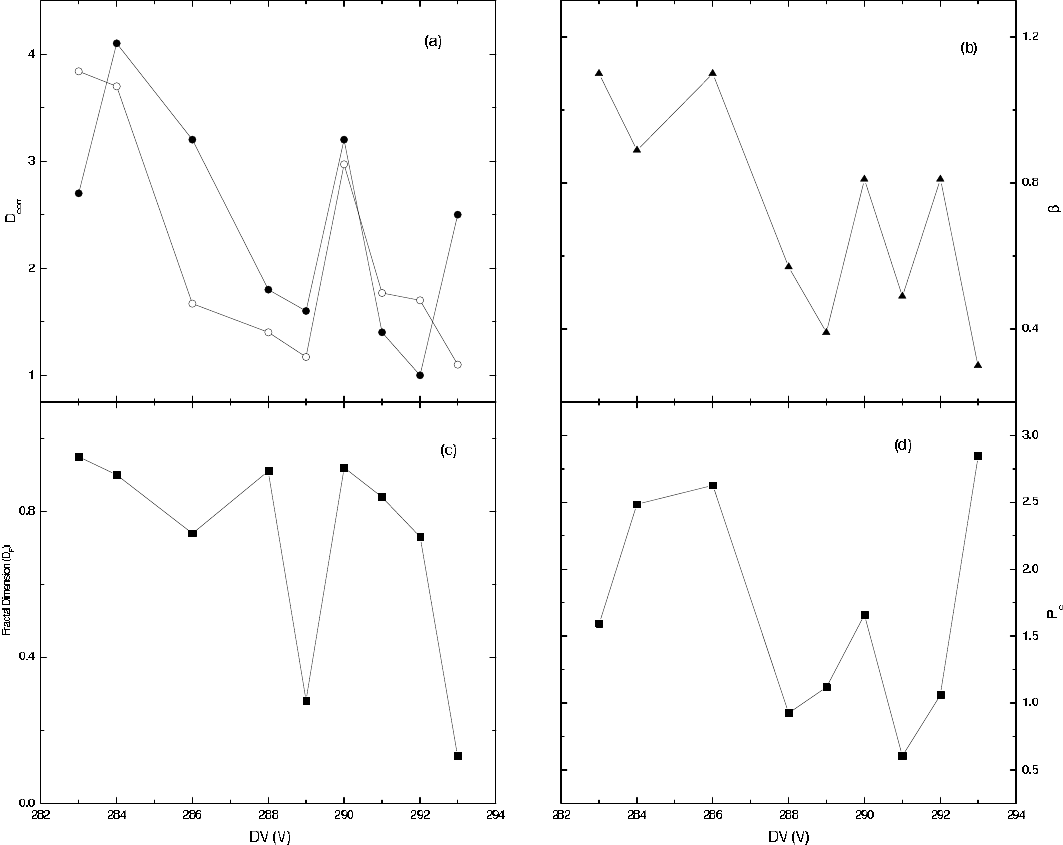}
\caption{(a) Correlation dimension ($D_{corr}$) obtained from the multi-fractal spectra
($-\bullet-$) and $D_{corr}$ from Grassberger-Procaccia algorithm (open
circle); (b) degree of multi-fractality ($\beta$); (c) fractal dimension ($D_F$); and
(d) complexity parameter ($P_c$) for different DV.} \label{fig:frac_corrbetapc}
\end{figure*}

We have estimated the multi-fractal spectra for all the nine signals
shown in Fig~\ref{fig2:contour}(left panel) at different DV. In
Fig~\ref{fig:mult}, black dots ($-\bullet-$) show the typical
multi-fractal spectra ($D(\alpha)$ vs $\alpha$) at 283V for the
initial stage of discharge  and uptriangle ($-\triangle-$) for the
fluctuations at 292 V just before the stable state. These plots
show that the signals at the initial stage of discharge is
multi-fractal in nature and tends to become a mono-fractal with
increase in DV (for quasi-periodic signal) as the width and height
of the spectrum have decreased considerably. For the periodic signals
[Figs~\ref{fig2:contour}(d, e and i)], we obtained a
mono-fractal behaviour.

The correlation dimension ($D_{corr}$) has been 
estimated from their multi-fractal spectra using the relation
$D_{q=2}=2\alpha_{q=2}-f(\alpha_{q=2})$~\cite{prb:Grussbach},
has been shown in Fig~\ref{fig:frac_corrbetapc}(a) by black dots
($-\bullet-$). It is seen that the $D_{corr}$ is high for the initial stage
of discharge and decreases with DV except for 290 V and 293 V. The
increase in $D_{corr}$ at 290 V is probably due to the appearance of a chaotic
feature at an intermediate stage of the periodic signal. 

For comparison, $D_{corr}$ using Grassberger-Procaccia algorithm~\cite{Chaos:jaman},  has been shown in the same plot (open circle) and both exhibit a good agreement. There are slight discrepancies observed at 283 V, 286 V and 293 V. The discrepancies at the first two values can be possibility a result of the presence of modes at significantly different scales. This would mean that since wavelet transform is based on the multi-resolution analysis of varying frequency and time resolutions at differing scales, the WTMM estimates joining maxima at different scales might incur errors in the estimated values of $D_{corr}$. However, the general trend of decrease in $D_{corr}$, from the two techniques, is quite similar in nature. The increase in $D_{corr}$ at 293 V may be due to nonlinear nature of the relaxation oscillations. The degree of multi-fractality ($\beta$)
defined as the difference between the maximum and minimum values of $\alpha$ and
which also statistically gives the range of scale
invariance~\cite{physicaA:sun} is shown in
Fig~\ref{fig:frac_corrbetapc}(c). Since $\beta$ is also a measure
of complexity, it is seen that the complexity
[Fig~\ref{fig:frac_corrbetapc}(a)] decreases with increase in DV.
Fractal dimensions, i.e., $f(\alpha_{q=0})$  of the singularity
support of the signals at different the DV have been shown in
Fig~\ref{fig:frac_corrbetapc}(c) and they decrease with increase
in DV. Finally we have estimated another complexity parameter
($P_c$) which is high for the complex
signal~\cite{NeuroImage:Shimizu}, and is defined as
$P_c=\frac{h_{max}}{f(\alpha(q=0))}H_{FWHM}$, where, $h_{max}$ is
the singularity exponent at the fractal dimension
($f(\alpha_{q=0})$) and $H_{FWHM}$ is the full width at half maxima
of the spectrum. Fig~\ref{fig:frac_corrbetapc}(d) shows the behavior of $P_c$
at different DV. At the initial stage of the discharge $P_c$ is high
and decreases at 288 V and 289 V, which corroborates  well with
estimated $D_{corr}$ [Fig~\ref{fig:frac_corrbetapc}(a)]. $P_c$
increases at 290 V where the signal is complex
[Fig~\ref{fig:frac_corrbetapc}(c)], and for the last case we have
high $P_c$ which may be due to the nonlinear effect  of the relaxation
oscillation [Fig~\ref{fig2:contour}(i)].

\section{Discussion and Conclusion}
\label{sec:conclusion} Glow discharges are simple systems but
exhibit exotic features depending on the configuration, discharge
parameters, etc. The complexities in the plasma dynamics arise
from many degrees of freedom like different sources of free
energy, various types of wave particle interaction and many other
instabilities. As the frequencies of the instabilities presented
in this paper, were within the ion plasma frequency and  the presence
of anode glow and relaxation oscillations have been attributed to
the presence of double layers, these instabilities could be
related to ion acoustic  waves in the presence of ionization
instabilities and also anode glow related double
layer~\cite{Chaos:jaman}. Such double layer associated ionization
instabilities have also been observed before ~\cite{jnlphysD:johnson}.
Different analysis tools have proved to be very  useful in
exploring the various possible linear and nonlinear features of
these instabilities.

Here, we have carried out wavelet based time-scale and
multi-fractal analysis of the floating potential fluctuations of
the glow discharge plasma for extracting some of the nonlinear
features.  From the contour plots of the CWT coefficients, we have
detected scale shift (or inveresly frequency shift) with time in
the glow discharge plasma.  Though theoretical prediction of such nonlinear
phenomenon has been reported in plasmas
~\cite{prl:Dawson,prl:Morales,pf1:vaclavik,pf2:ishihara}, there are very few experimental evidences  on this~\cite{pop:Koepke}, and we have identified this phenomenon using wavelet techniques. The scale shifts can occur if  the coherent modes like the ion acoustic wave or a double layer
propagates through an ion acoustic
turbulence~\cite{pf1:vaclavik,pf2:ishihara}. The ridge patterns of
the plasma signals based on CWT decomposition  reveal the chaotic
features of the system and these finding agree well with our
earlier analysis using the nonlinear time series
analysis~\cite{Chaos:jaman}. The estimated correlation dimension,
fractal dimension, and complexity parameter from multifractal
spectra are also in close agreement  with the nonlinear
analysis~\cite{Chaos:jaman}. With increase in the DV, the multi-fractal
behaviour changes to a monofractal as the irregular oscillations
become periodic with increase in the DV.

Finally, we feel, the analysis of chaotic data using the contours,
ridge plots and multi-fractal spectra, based on the CWT decomposition,
are very useful because these  do not require a long data length.
However  one should be cautious about the choice of the
appropriate wavelet. Whereas, the estimation of the correlation
dimension and Lyapunov exponents estimated from the nonlinear time series
analysis require a long data length, and these estimations depend also
on the proper choice of the embedding dimension, time lag, etc.

\section*{acknowledgement}
We acknowledge the constant support and encouragement from the Director, SINP.  We would also like to  thank our colleagues of the plasma physics division for their help during the experiments. One of the authors (MN) would like to acknowledge the support of Amit Apte at TIFR.

\end{document}